\documentclass{ws-ijmpa}
\usepackage{cite}
\usepackage{amsfonts}
\usepackage{amsmath}
\usepackage{amssymb}

\newcommand{\be}{\begin{equation}}\newcommand{\ee}{\end{equation}}
\newcommand{\bea}{\begin{eqnarray}}\newcommand{\eea}{\end{eqnarray}}
\newcommand{\beaa}{\begin{eqnarray}}\newcommand{\eeaa}{\end{eqnarray}}
\newcommand{\ba}{\begin{array}}\newcommand{\ea}{\end{array}}
\newcommand{\bit}{\begin{itemize}}\newcommand{\eit}{\end{itemize}}
\newcommand{\ben}{\begin{enumerate}}\newcommand{\een}{\end{enumerate}}

\def\lab{\label}

\begin{document}

\title{Exploring Properties of Dark and Visible Mass
Distribution on Different Scales in the Universe}

\author{Yuriy Mishchenko and Chueng-Ryong Ji}
\address{Physics Department \\
 North Carolina State University, Raleigh, NC }
\maketitle

\begin{abstract}
In this short note we discuss recent observation of 
linear correlation on log-log scale between distribution of 
dark and visible mass
in gravitationally bound systems. 
The coefficient of such correlation
appears to be essentially the same for various
systems of dramatically different scales such as
spiral galaxies of different luminosities and
galaxy clusters.
We briefly touch possible interpretations
of this observation and implications for
the mass of dark matter particle.
\end{abstract}

Presence in the Universe of large amounts of dark (non-luminous) matter has
been one of the central puzzles in cosmology since
the beginning of last century. Starting from 1920th, it had been
observed that Rotation Curves (RC)
in spiral galaxies
measured up to twice 
the extent of the galaxy stellar disk show yet no
Keplerian fall-off. 
These observations provided first indications 
that spiral
galaxies are submerged into vast, isothermal,
almost spherical dark halos.
More insight had been obtained in the recent years
using
phenomenon of gravitational lensing
which spectacularly showed
presence of huge amounts of nonluminous
matter in the galaxy clusters.
These days, thanks to
extensive observations and computer simulations, 
we gained a better
understanding of dark matter and its role in the Universe evolution. 
Given observed smoothness of dark mass distribution,
large mass-to-light ratio and failure of earth-based
experiments to get reliable dark-matter events,
it is believed that dark matter is made of 
nonbaryonic heavy weakly interacting particles.
A number of candidates in various extensions of 
the Standard Model had been put forward 
but any of them are yet to be confirmed
in experiments.
Recently an interesting property
of dark and visible mass distribution in gravitationally bound
systems has been pointed out, which may be ultimately related to
microscopic properties of dark matter. Specifically,
based on the study of rotation curves in a large sample 
of spiral galaxies by Persic,  Salucci and Stel \cite{perssalu}
and the analysis of mass distribution in galaxy
cluster CL0024 
by Tyson, Kochanski and Del'Antonio \cite{tysokoch}, we observed
that the distribution of dark and visible mass
in these systems is well consistent with linearly
correlated on log-log scale with
what appears to be a universal
correlation coefficient $\kappa \approx 3.5 \pm 1.5$
\cite{mishji}.

\begin{figure}
\begin{minipage}[c]{0.3\hsize}
\centerline{\psfig{file=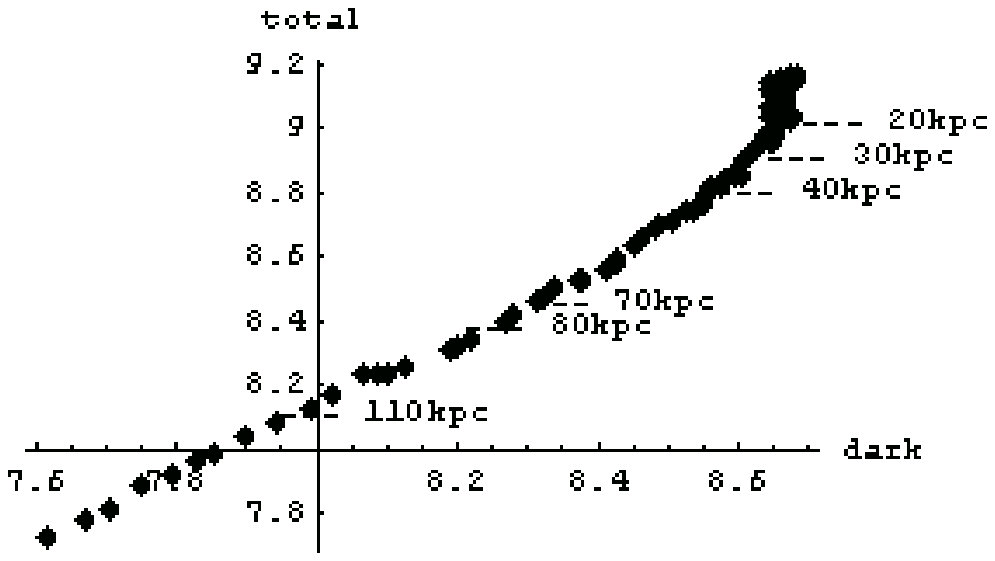,width=\hsize}}
\end{minipage}
\begin{minipage}[c]{0.3\hsize}
\centerline{\psfig{file=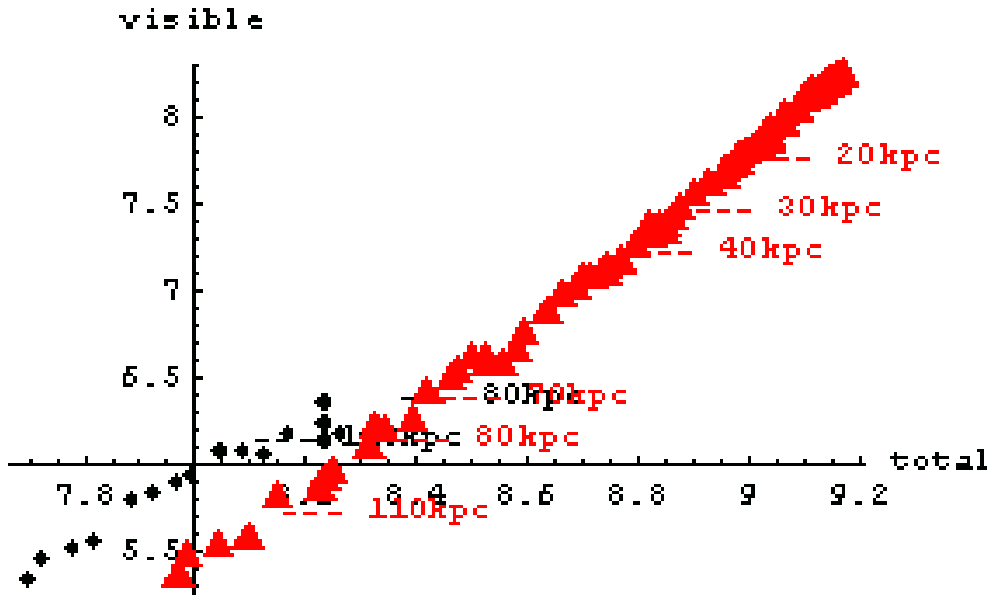,width=\hsize}}
\end{minipage}
\begin{minipage}[c]{0.3\hsize}
\centerline{\psfig{file=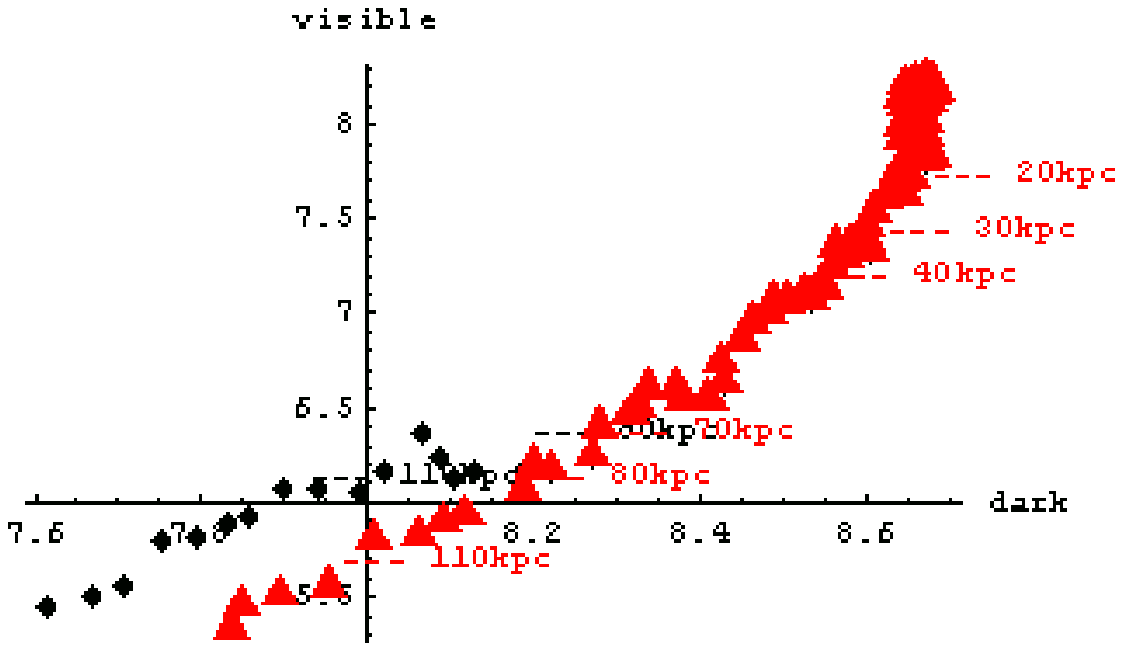,width=\hsize}}
\end{minipage}
\caption{}
\label{fig1}
\end{figure}

In 1998, Tyson {\it et al}  \cite{tysokoch} carried out analysis of
the Hubble Space Telescope 
images of strong gravitational lensing in
galaxy cluster CL0024+1654. This cluster is one of the
examples of strong gravitational lensing in which
a number of images of a background galaxy with distinctive
color is formed. 
Detailed distribution of
mass and azimuth-averaged
distributions of total and visible mass in the cluster had been obtained.
The primary conclusion of this study was the
failure of
Cold Dark Matter simulations in description
of the observed flat-core mass profile.
We have reexamined the mass profiles
presented in \cite{tysokoch}
with the emphasis of search for relations between
dark and visible components
and observed that if
any two mass profiles are plotted one vs another
on log-log scale the data points form a 
straight line. Deviations
from this linear correlation are only observed
in small region in the center of the cluster.
As is shown in Figs.(\ref{fig1}a), e.g.,
very good linear correlation is present
between log of total and log of dark mass
distributions with correlation coefficient
of about $\kappa_{td}\approx 1.15-1.9$
for a region of well over 100kpc.
The same effect is observed for any two
profiles in the data, as can be seen
in Fig.(\ref{fig1}b) and (\ref{fig1}c). 
The correlation coefficients in these cases 
were found to be for visible-total mass
$\kappa_{vt}\approx 1.4-2.25$ and 
for visible-dark mass
$\kappa_{vd}\approx 2.10-4.40$.
Such feature in the
galaxy cluster CL0024 is rather interesting
and may indicate approximate large scale
thermal equilibrium in
both visible and dark components in the cluster.
In that case,
\be\lab{E1.a}
\ln \rho_v = \frac {\mu_v} {T_v} \Phi \sim \ln \rho_d = \frac {\mu_d}{T_d}\Phi.
\ee
Still, to make this result more significant 
we would like to see more examples of this behavior
in other systems.
While we do not know of  comparable studies
of galaxy clusters, we note that there is yet another observable
heavily affected by the presence of dark matter, i.e.
RC in spiral galaxies.

In Persic {\it et al} analysis \cite{perssalu}
of a sample of over 1000 spiral galaxies  had
shown that RCs grouped by galaxy luminosity
and normalized to dimensionless units
via optical radius $r_{opt}$
(i.e. extent of the stellar disk)
and
rotation velocity at the optical radius
$v_{opt}$ follow universal profiles
called by the authors Universal Rotation Curves (URCs).
In \cite{perssalu}, URCs had been fit using simple
mass model to within their rms.
In this fit the stellar disk is described
as a thin disk with
the surface mass density
$\rho_{v}(r)\sim exp(-3.2 r/r_{opt}).$
This can be motivated with the notion
that visible mass distribution should
follow the surface brightness in the
stellar disk which decays
exponentially as is well known
from the stellar-disk photometry.
The gravitational potential from exponentially
thin disk is known and contribution
from visible mass into RC $v_v^2$ in this model
can be obtained analytically.
In \cite{perssalu}, this distribution was modeled
for convenience with a good analytical fit
within few percents.
The contribution from the dark halo  
was parametrized in the form
\be\lab{E3}
v_{d}^2 \approx (1-\beta)(1+\alpha^2) \frac
{x^2}{x^2+\alpha^2}.
\ee
Here $\alpha$ and $\beta$ are the fit
parameters describing extent and amount
of the dark halo in the galaxy.
RC is described in this way as
$v^2_{URC}=v^2_v+v^2_d$.

Guided by the idea of log-log-linear correlation,
we investigated the question if
URCs can be reproduced and to what accuracy
with the dark mass distribution
\be\lab{E4}
\rho_d \sim \exp( -a r/r_{opt} ).
\ee
For all absolute luminosities that Persic {\it et al} 
applied their model (i.e. $-23 < M < -19$) we found that 
a perfect fit can be obtained to the URC
with spherical dark halo distributed according
to Eq.(\ref{E4}).
The best fit coefficient $a$ varies quite moderately
with galaxy luminosity $M$ and is in
the range of $\kappa_{vd} \approx 2.5-4.11$.

The coincidence of this estimate with
$\kappa_{vd}\approx 2.1-4.4$ in CL0024 is striking.
Log-log-linear correlation, thus, may be seen
as an indication of approximate large scale
thermal equilibrium
in visible and dark components:
$\kappa \approx T_d/T_v\cdot \mu_v /\mu_d$.
$\kappa_{vd}\approx const \approx 3 $ can
be understood as 
$T_v/T_d \approx 1$ so that 
$\kappa_{vd} \approx \mu_v / \mu_d$
and $\mu_d \approx 200-1000MeV$.
While these phenomenological conclusions
are quite coherent, it is
hard 
to understand how such thermal equilibrium
can be reached
given our current views about dark matter. 
One may note that spiral galaxy is
not a virialized object, unlike galaxy cluster.
Thus, significance and existence of such correlation can not be easily 
understood. 
Also, one would not expect $\kappa_{vd}$ to be the same
in different systems. 
More reasonable expectations would be
either $\kappa \approx 1$ in galaxy clusters
(primarily due to gravitational heating
in which case
$T_{v,d} \approx \mu_{v,d} \Delta \Phi$
and $\kappa \approx 1$)
or $\kappa >> 1$ in spiral galaxies (cooling
of visible component is strong while
dark matter is noninteracting and nondissipative).
In this respect, examining
alternative interpretations of the
log-log-linear correlation is important.
Among such alternatives is 
some unknown underlying feature
of structure formation,
thermalization due to
purely
gravitational interaction, 
or even relations of dark matter to 
QCD condensates due to $\mu_d \approx \Lambda_{QCD}$.
Finally, given a large number of 
uncertainties in any astrophysical study of
this sort, e.g. visible-dark
separation in RC
or uncertainties in inferring visible matter amount, 
a larger number of 
observations of log-log-linear correlation
is desirable.
In this sense a large survey of
gravitational lensing in galaxy clusters with
emphasis on dark vs. visible matter
distribution would be most beneficial.

This work was supported in part by a grant from the U.S. Department of
Energy (DE-FG02-96ER 40947). The National Energy Research Scientific
Computer Center is also acknowledged for the grant of computing time.

\end{document}